\documentclass[conference]{IEEEtran}
\IEEEoverridecommandlockouts
\usepackage{cite}
\usepackage{amsmath,amssymb,amsfonts}
\usepackage{graphicx}
\usepackage{textcomp}
\usepackage{xcolor}

\usepackage{comment}
\usepackage{hyperref}
\hypersetup{colorlinks=true,linkcolor=black,urlcolor=black,citecolor=black} %
\usepackage{enumitem}
\usepackage{flushend}  %

\def\BibTeX{{\rm B\kern-.05em{\sc i\kern-.025em b}\kern-.08em
    T\kern-.1667em\lower.7ex\hbox{E}\kern-.125emX}}

\newcommand{\ie}{{i.e. }}
\newcommand{\eg}{{e.g. }}
\newcommand{\etal}{{et al. }}

\begin{document}

\begin{titlepage}
\quad\\[1cm]
\makeatother
	{\Huge IEEE Copyright Notice}\\[0.5cm]
	{\large \copyright \ 2024 IEEE. Personal use of this material is permitted. Permission from IEEE must be obtained for all other uses, in any current or future media, including reprinting/republishing this material for advertising or promotional purposes, creating new collective works, for resale or redistribution to servers or lists, or reuse of any copyrighted component of this work in other works. \\[0.5cm]}    
    {\large Cite as:\\[0.1cm]}
    {\large D. Margaria, A. Carelli and A. Vesco, “Building Trust in Data for IoT Systems,” 
    in \textit{Proceedings of IEEE 24\textsuperscript{th} International Symposium on Cluster, Cloud and Internet Computing Workshops}, Philadelphia, PA, USA, 2024, pp. 1-8, doi: 10.1109/CCGridW63211.2024.00006.}
\end{titlepage}

\title{Building Trust in Data for IoT Systems
\thanks{%
This work has been developed within the MASTERMINE project (European Mining in the Green and Digital Era, \url{https://www.mastermine-project.eu/}), funded by the European Union under the Horizon Europe framework programme [GA 101091895].}}

\author{\IEEEauthorblockN{Davide Margaria\href{https://orcid.org/0000-0002-5275-2138}{\includegraphics[height=\fontcharht\font`B]{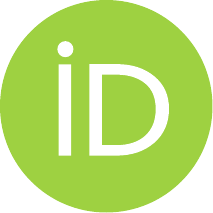}}, 
Alberto Carelli\href{https://orcid.org/0000-0003-2392-5463}{\includegraphics[height=\fontcharht\font`B]{Figures/logo-orcid.pdf}}, 
Andrea Vesco\href{https://orcid.org/0000-0001-7431-6655}{\includegraphics[height=\fontcharht\font`B]{Figures/logo-orcid.pdf}}} %
\IEEEauthorblockA{%
\textit{LINKS Foundation - Cybersecurity Research Group - 
Torino 10138, Italy} %
}}

\maketitle

\begin{abstract}
Nowadays, Internet of Things platforms are being deployed in a wide range of application domains. 
Some of these include use cases with security requirements, where the data generated by an IoT node is the basis for making safety-critical or liability-critical decisions at system level.  
The challenge is to develop a solution for data exchange while proving and verifying the authenticity of the data from end-to-end.
In line with this objective, this paper proposes a novel solution with the proper protocols to provide Trust in Data, making use of two Roots of Trust that are the IOTA Distributed Ledger Technology and the Trusted Platform Module.
The paper presents the design of the proposed solution and discusses the key design aspects and relevant trade-offs.
The paper concludes with a Proof-of-Concept implementation and an experimental evaluation to confirm its feasibility and to assess the achievable performance.
\end{abstract}


\section{Introduction}
\label{sect:Intro}

\label{sect:context}

Internet of Things (IoT) nodes are now being deployed in a plethora of different connected systems and are moving towards an ecosystem phase.
Attacks based on malicious data ingestion, data manipulation, eavesdropping on contextual information indicative or derived from IoT data represent a serious problem especially in specific scenarios like autonomous cars in cooperative intelligent transport systems %
\cite{Amanullah2024} and critical infrastructures \cite{Sisinni2022}, 
potentially resulting in catastrophic incidents.
In these scenarios, safety-critical or liability-critical decisions are made on the basis of data generated or collected by IoT nodes, hence the need for \textit{Trust in Data} \cite{Guo2021, Ren2022, Mo2020}.
In this context, there are already some examples of solutions that take advantage of Distributed Ledger Technology (DLT) \cite{DLTs} to preserve the integrity of the data stored in the distributed ledger against potential malicious manipulation \cite{Alam2023, Alajlan2023, Bikos2022, Montanaro2023, Khelifi2018, Wang2020}. In addition, the communication between the IoT nodes and the DLT gateway node is often protected by the Transport Layer Security (TLS) protocol \cite{TLS}. 
It must be noted that the security properties provided by the DLT (\ie immutability, hence integrity of the data stored) and secure communications (\ie authentication, data integrity and confidentiality in transit) are \textit{not enough} to take critical decisions on data generated by IoT nodes. 
The combination of the two solutions basically provides cryptographic protection of the data transmission from the source to the final destination, but they do not consider the software (SW) integrity of the originating IoT node \cite{Margaria2021, Sisinni2022}. If the SW stack of the IoT node is compromised, the data can be manipulated at the origin before being transmitted over the secure channel and therefore cannot be fully trusted.

\label{sect:contrib}

Aiming to solve the problem highlighted above, this paper proposes a novel solution to ensure the trust in data, effectively providing end-to-end protection of the entire chain from data generation to final data consumption for decision-making.
The proposed solution, called \textit{Trusted Data over the Tangle} (TDT), is based on two Root of Trust (RoT): 
\begin{enumerate}
    \item the IOTA Tangle as the RoT for data \cite{Tangle}, 
    \item the Trusted Platform Module (TPM) as the RoT for SW integrity measures and reporting \cite{tpm20, TPMbook}.
\end{enumerate}

Excluding constraint IoT, today IoT nodes in connected systems can easily be equipped with a TPM, \eg Infineon OPTIGA{\texttrademark} TPM SLI 9670 board for Raspberry Pi{\textsuperscript{\textregistered}} 4 \cite{infineonTPM}. 
Hence, this paper assumes that each critical IoT node in the system is equipped with a hardware TPM enabling the verification of SW integrity of the data source, while the IOTA Tangle ensures data immutability for near real-time and non-real-time data consumption. 

The contribution of this paper is threefold: (\emph{i}) proposes the TDT concept and design for IoT connected systems that need to produce, exchange and consume trusted data to enable critical decision-making, (\emph{ii}) defines the protocols ruling the interactions between the involved agents and highlights the relevant trade-offs, and (\emph{iii}) presents a Proof-of-Concept (PoC) implementation and reports promising experimental results that confirm the feasibility of the TDT solution.

\section{Related Works}
\label{sect:rw}

\subsection{Existing Approaches to Provide Trust in IoT Data}
\label{sect:ex_approaches}

Several scientific papers have proposed solutions to evaluate and enhance the trustworthiness of IoT data, based on different assumptions and theoretical foundations \cite{Guo2021, Ren2022, Mo2020, DLTs, Alam2023, Alajlan2023, Bikos2022, Montanaro2023, Khelifi2018, Wang2020}. 

Guo~\etal\cite{Guo2021} introduce the Lightweight Verifiable Trust based Data Collection (LVT-DC), an approach for sensor to cloud systems. The work considers a scenario with many IoT nodes in a smart city and a set of mobile vehicles that collect data from those nodes and report data to the cloud to enable different applications.
The LVT-DC solution evaluates the trustworthiness of the mobile vehicles, based on direct trust computation and trust inference computation. 
The LVT-DC solution considers the IoT node trustworthy by design and focuses on estimating  the reliability of unknown data collectors.
Instead, no IoT node is considered \textit{a priori} trusted in our work, where the TDT solution takes advantage of a TPM to verify the integrity of the entire SW stack of each IoT node and its configuration, according to Trusted Computing (TC) principles \cite{TPMbook}.

Ren~\etal\cite{Ren2022} propose the Privacy-protected Intelligent Crowdsourcing scheme based on Reinforcement Learning (PICRL).
This solution assesses the trustworthiness of unknown data sources and aims to optimise the utility of crowdsourcing applications by simultaneously considering the data amount, data quality, and cost.
Notably, the trustworthiness calculation of IoT nodes is \textit{probabilistic} due to the inherent characteristics of the system.
Instead, our work performs a deterministic evaluation of the trustworthiness of the IoT nodes by assessing their SW integrity with TC techniques.

Mo~\etal\cite{Mo2020} present the Active and Verifiable Trust Evaluation (AVTE), an approach to enhance the reliability of edge computing applications, whose decisions are based on data forwarded by sensing devices that are self-organised in IoT networks. 
The AVTE solution evaluates the credibility of IoT nodes as trustworthy forwarding nodes, to ensure reliable and low cost data collection for edge computing. 
Again, this approach does not assess the SW integrity of IoT nodes, nor does it cover attacks aimed at compromising the data sources.
Moreover, the solutions in \cite{Guo2021, Ren2022, Mo2020} do not use a TPM and/or DLTs, which are the key building blocks of our work.

Concerning the DLT-based solutions \cite{DLTs}, the recent scientific literature includes relevant examples of DLT adoption in different categories of IoT systems \cite{Alam2023, Alajlan2023, Bikos2022, Montanaro2023, Khelifi2018, Wang2020}.

Among the others, it is worth to mention the works of 
Alam~\etal\cite{Alam2023}, Alajlan~\etal\cite{Alajlan2023}, and  
Bikos~\etal\cite{Bikos2022}, that present comprehensive literature surveys of DLT-based solutions suitable for IoT systems, with detailed analyses of current trends, cybersecurity requirements, vulnerabilities, future challenges and possible areas for improvement. 

Bikos~\etal\cite{Bikos2022} highlight the IOTA Tangle as a promising distributed ledger capable of addressing both the security and performance challenges of traditional IoT applications.

Montanaro~\etal\cite{Montanaro2023} analyse the benefits and the performance of the simultaneous use of an off-chain database, to store actual IoT data, and a DLT, to guarantee the immutability of such data and other relevant properties.

Khelifi~\etal\cite{Khelifi2018} present a reputation-based blockchain mechanism tailored to a vehicular network scenario and suitable to a Named Data Networking (NDN) architecture. 
This solution secures the in-network caching and enforces the trust between cache stores and consumer vehicles by assigning each cache store a reputation value, that is increased/decreased based on the served content.

Wang~\etal\cite{Wang2020} propose a mutual authentication and a key agreement protocol that takes advantage of a DLT to verify the IoT data sources and ensure data integrity in a smart grid with edge computing scenario. 

All the works in \cite{Alam2023, Alajlan2023, Bikos2022, Montanaro2023, Khelifi2018, Wang2020} take advantage of DLTs to ensure the integrity of IoT data stored. However, they do not %
consider the SW integrity of the IoT nodes that actually generate the data. This is the main shortcoming of the existing solutions.

With this work, we are exploring the feasibility and benefits of combining the use of DLT and hardware TPM to build trust in data, thereby enhancing the security and reliability of IoT-connected systems.

\subsection{Enabling Building Blocks for the Proposed Solution}
\label{sect:blocks}

The proposed TDT solution takes advantage of the following building blocks:
\begin{itemize}
    \item \textit{IOTA Tangle} \cite{Tangle}, a DLT designed primarily for use in IoT environments and suitable for anchoring data while taking advantage of its data immutability feature;
    \item \textit{Wrapped Authenticated Messages} (WAM) \cite{WAM}, a novel open-source cryptographic protocol for structuring data on the IOTA Tangle and facilitating their consumption in near real-time and non-real-time;
    \item \emph{Trusted Platform Module} 
    \cite{tpm20}, a tamper-resistant cryptographic hardware at the core of TC principles \cite{TPMbook} on top of which TDT builds its own SW integrity architecture;
    \item \emph{Remote Attestation} (RA) protocol \cite{TAP, rfc9334} to challenge and verify the SW integrity of IoT nodes at run-time.
\end{itemize}

The IOTA Tangle \cite{Tangle} is a permissionless public DLT available for anchoring data in a feeless manner. WAM is an open-source application level protocol~\cite{WAM} to structure a stream of data on the IOTA Tangle. To the best of our knowledge, WAM is the only open-source implementation in C language available for IoT nodes. It organises data into concatenated messages that form a unidirectional logical channel on the IOTA Tangle. Optionally, WAM provides the function to encrypt/decrypt data. 
The logical channel allows any node in the system to subscribe to a WAM channel of another node and retrieve the data in near real-time or rapidly retrieve the full chain of data (or a portion of it) in non-real-time, while verifying the authenticity of the retrieved data. 
The WAM protocol signs each message and provides the functionality to verify the authenticity of each message at the subscriber end, removing this complexity from the applications.  
Moreover, the  WAM protocol ensures the channel ownership property. Only the owner of a WAM channel can publish data on it, while any malicious attempt by an adversary to take control of the channel or to insert false data can be easily detected by any node subscribed to it during the verification of the data retrieved \cite{WAM}. 

At this point of the paper, it should be agreed that the TLS protocol protects the data in transit from the IoT node to the IOTA Tangle, the WAM protocol ensures the authenticity of the data published by the IoT node in its WAM channel stored in the IOTA Tangle, and the IOTA Tangle ensures the integrity of the data thanks to its immutability feature (\ie authenticity, integrity and \emph{optionally} confidentiality are provided from end-to-end). The remaining security gap is the possibility for an adversary to tamper with the SW stack of the IoT node and modify the data at source. For this reason, TDT adds SW integrity verification using a hardware TPM and a Remote Attestation protocol.

\begin{figure}[tb] %
    \centerline{\includegraphics[width=\columnwidth]{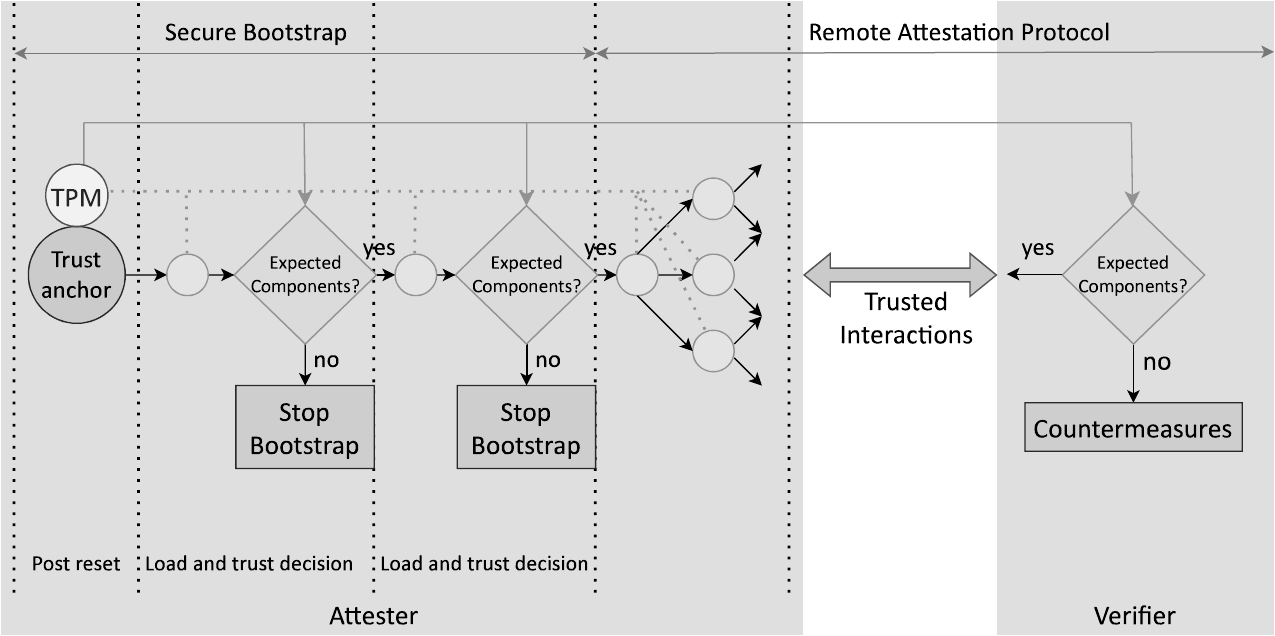}} %
    \caption{Software integrity architecture.}
    \label{fig:RA}
\end{figure}

The SW integrity architecture is depicted in Fig.~\ref{fig:RA}. The IoT node implementing the TC principles on top of a hardware TPM \cite{tpm20} is called Attester.  
The identification of the SW components running on the Attester and the verification of their integrity is performed by means of a hierarchy of trust rooted into a trust anchor (\ie typically a small portion of code into the processor executed at the very beginning of the boot process). 
The architecture implies different trust decisions to decide whether the Attester can be considered trusted or not. Such decisions can be taken by the Attester itself during the bootstrap (\eg before running the bootloader and before running the OS Kernel) to build the so-called secure bootstrap. Moreover, the decision can also be taken by a Verifier running the RA protocol \cite{TAP, rfc9334}. 
The RA is a challenge/response protocol in which the Verifier is able to determine the trustworthiness level of the Attester against a list of golden values (\ie the hash values of the software components running on the Attester). 
The adoption of a RA protocol enables periodical verification of the Attester integrity at run-time, so it adds security to the sole verification at boot time. 
With reference to Fig.~\ref{fig:RA}, the first two trust decisions are typically made by the processor based on the signature of the bootloader and of the OS Kernel. Upon full loading of the OS, the third periodical decisions are made by the Verifier through remote attestation.
These decisions are made based on (\emph{i}) the values of the Platform Configuration Registers (PCRs) inside the hardware TPM that accumulate the digest of the SW components running on the Attester, and (\emph{ii}) on the values of the Integrity Measurement Architecture (IMA) Measurement Log (ML)~\cite{imapaper}. %
For more details on these aspects, interested readers can also refer to \cite{TPMbook, tpm20}. 

The RA protocol is addressed in standards and specifications, for instance the Trusted Attestation Protocol (TAP) Information Model defined by the Trusted Computing Group (TCG) \cite{TAP} and the Remote ATtestation procedureS (RATS) architecture by the Internet Engineering Task Force (IETF) \cite{rfc9334}.
Moreover, several RA solutions have already been proposed and implemented in different application domains and use cases, for instance critical infrastructures \cite{Sisinni2022}, Industry 4.0 \cite{Margaria2021}, or time distribution networks \cite{Berbecaru2023}.

\section{Trusted Data over the Tangle (TDT)}
\label{sect:solution}

\subsection{General Principles and Agents} 

The proposed TDT solution, depicted in Fig.~\ref{fig:Agents}, involves the following agents: 

\begin{figure}[t]
    \centerline{\includegraphics[width=\columnwidth]{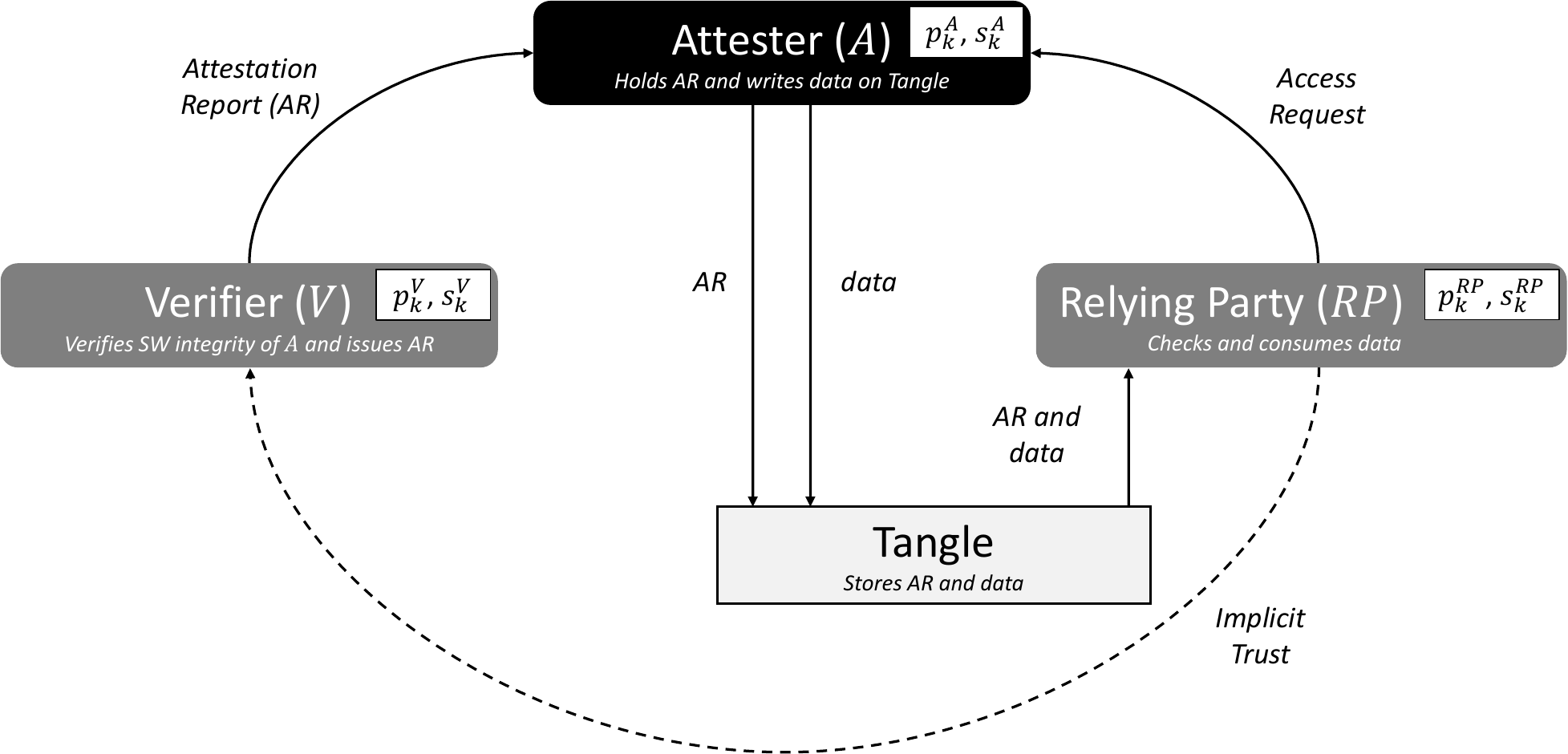}}
    \vspace{1mm}
    \caption{Agent involved and their roles.}
    \label{fig:Agents}
\end{figure}

\begin{description} 
	\item[{Verifier ($V$)}] -- a specialised agent in charge of enforcing the trustworthiness level of IoT nodes (\ie Attester) in the system. 
    It verifies the digital identity and the SW integrity of the nodes by means of a RA protocol. The Verifier generates, signs, and sends back an \textit{Attestation Report} (AR) to the IoT nodes.
	\item[{Attester ($A$)}] -- any IoT node in the system that generates data that other IoT nodes in the system consume. 
    It is equipped with a hardware TPM and it initiates the RA protocol by contacting the Verifier asking for an AR.	
	Upon receiving a valid AR, the Attester starts writing the data into its WAM channel, inserting with a configurable periodicity a fresh AR in the stream. 
    To preserve the confidentiality of the data, the Attester can optionally encrypt them. For instance, the Attester can take advantage of Authenticated Encryption with Associated Data (AEAD) through a specific security layer of WAM protocol \cite{WAM}.
	\item[{Relying Party ($RP$)}] -- any IoT node in the system that consumes the data of an Attester through the Tangle. 
    After subscription to the selected WAM channel, the Relying Party performs the appropriate verification (\ie on the identity and SW integrity of the Attester) before trusting and consuming the data.
 	Note that the Relying Party carries out an \emph{a posteriori} check on the AR and data. 
	\item[{Tangle}] -- is the RoT for the data generated by the Attesters in the IoT system.
\end{description}

For the sake of simplicity but without losing generality, Fig.~\ref{fig:Agents} shows a single agent for each role. In principle, it is possible to generalise this basic scheme to more complex use cases, based on multiple Verifiers, Attesters, and/or Relying Parties. Moreover, an IoT node can play both the role of Attester and Relying Party. 

It must be noted that the proposed system is compliant with the RATS architecture defined in the RFC 9334 \cite{rfc9334}. In detail, the Verifier, Attester, and Relying Party agents in Fig.~\ref{fig:Agents} correspond to the roles formally defined in \cite{rfc9334}. Moreover, the data flows are consistent with the topological pattern named as \textit{Passport Model} in \cite{rfc9334}, due to its resemblance to how a national authority issues a passport to a citizen and how an immigration desk checks that passport.
In this analogy, the Attester corresponds to a citizen that conveys a passport application and identifying information (\ie an evidence) to a passport-issuing agency, that is the Verifier.
The Verifier compares the evidence against its appraisal policy and, if the evidence passes the appraisal policy, the Verifier  gives back a passport (\ie AR).
The Attester treats the passport as opaque data and appropriately stores it.
The Attester can now present (\ie write on its WAM channel on the Tangle) the passport (\ie AR) and possibly additional claims (\ie data) to an airport immigration desk (\ie Relying Party).
The immigration desk then compares this information against its own appraisal policy to make an authorisation decision (\ie to consume or discard the AR and data). 

\subsection{Assumptions and Notation} 
\label{sect:assumptions}

The TDT description assumes the use of any public-key signature scheme supported by the TPM (\eg RSA, ECDSA, ECDAA) \cite{TCGalgoRegistry}.

A proper \textit{provisioning} phase is assumed for all the IoT nodes in the system. 
This phase is carried out before deployment of IoT nodes on the field and it consists of the typical initial configuration procedure in a secure environment. %
It includes the generation of a database of \textit{golden values}, that consists in the computation of the digests for all the relevant SW components on the IoT node that will be verified during the RA protocol,
and the execution of a proper key generation procedure.  
In detail, the Verifier, Attester, and Relying Party execute a key generation algorithm to initialise the asymmetric key pairs associated to their identities, according to the selected signature scheme.

Aiming to unambiguously identify the owner of each public key ($p_k$) and private key ($s_k$), the next sections adopt the following notation: 

\begin{itemize}
		\item the Verifier key pair $(p^V_k, s^V_k)$;
		\item the Attester key pair $(p^A_k, s^A_k)$;
		\item the Relying Party key pair $(p^{RP}_k, s^{RP}_k)$.
\end{itemize}

It is worth highlighting that the TPM in the Attester node generates the key pair $(p^A_k, s^A_k)$ according to an appropriate key hierarchy.
Such asymmetric key pair is conventionally denoted as the \emph{Attestation Key} (AK) \cite{TPMbook, tpm20}.
It is a non-migratable key pair, since its private part $s^A_k$ will never leave the TPM, while $p^A_k$ is publicly available.
In principle, the AK is used to sign and verify the TPM Quote during the RA protocol and the WAM messages generated by the Attester, ensuring a hardware-rooted authenticity.
The TPM Quote consists in the signed list of PCR values stored in the TPM.
The AK is protected by means of the so-called \emph{Endorsement Key} (EK),
that corresponds to another non-migratable asymmetric key pair, derived from a seed statistically unique for the TPM.
The EK is certified by the TPM manufacturer and represents the root of the chain of trust for any AK.
The provisioning phase terminates with the distribution and the secure installation of all the public key certificates in all agents of the TDT system.

\subsection{The Protocols}
\label{sect:protocols} 

\begin{figure}[th]
    \centerline{\includegraphics[width=\columnwidth]{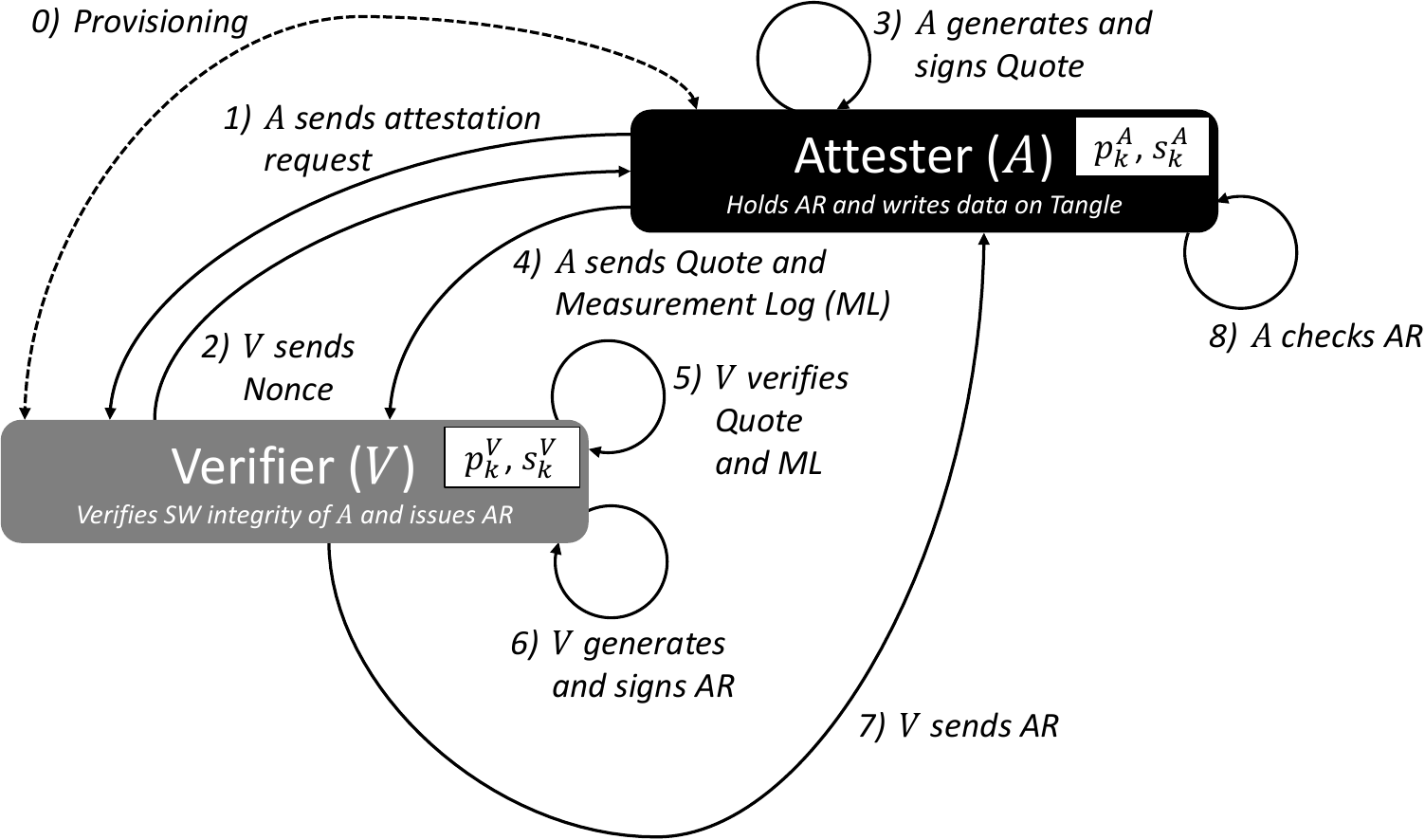}}
    \caption{The Remote Attestation protocol.} %
    \label{fig:RAprotocol}
\end{figure}

\begin{figure}[th]
    \centerline{\includegraphics[width=0.9\columnwidth]{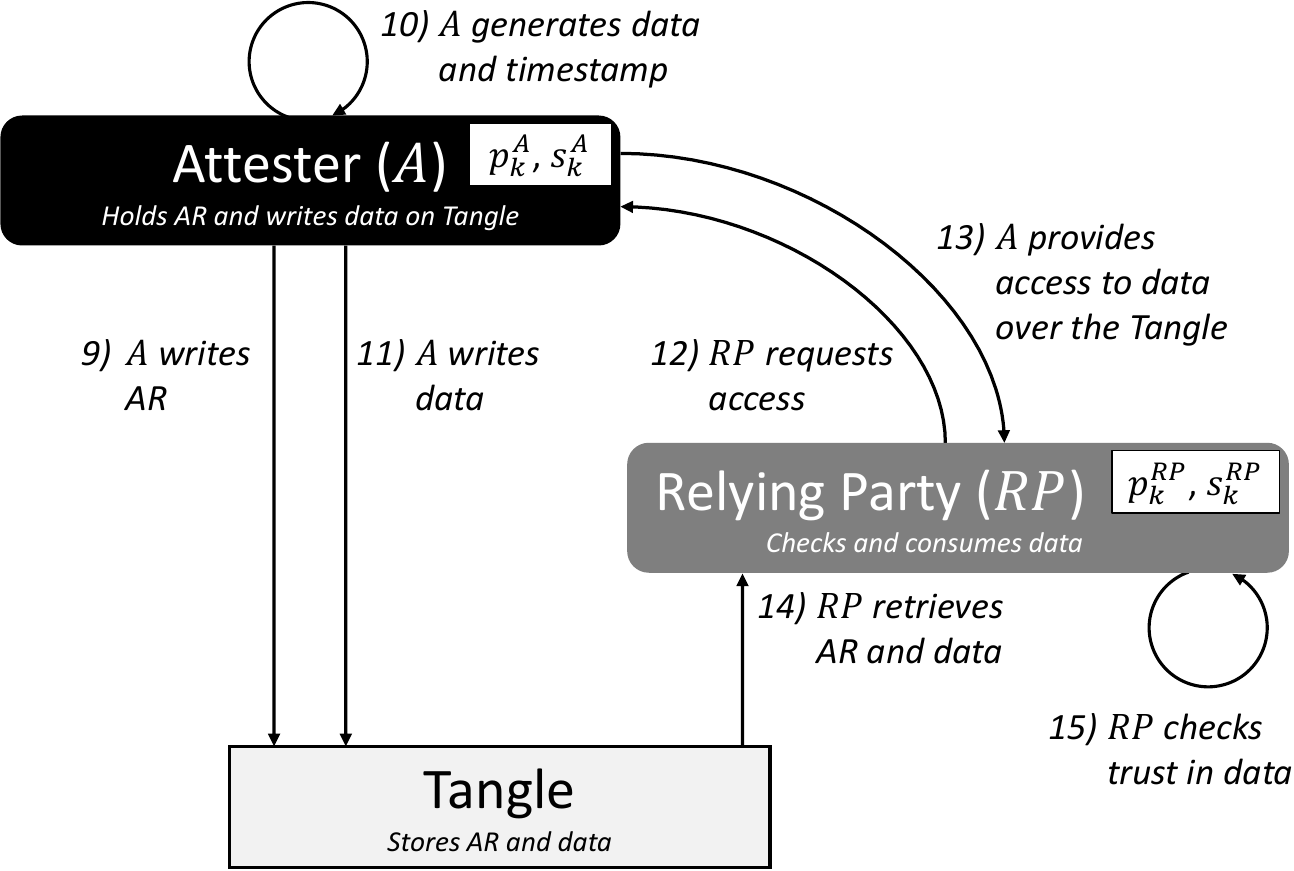}}
    \caption{The Data Exchange protocol.} 
    \label{fig:TangleProtocol}
\end{figure}

The TDT solution comprises two protocols that take place between different agents at different times: (\textit{i}) the RA protocol, between $A$ and $V$, and (\textit{ii}) the Data Exchange protocol, between $A$ and a $RP$ through the Tangle, as shown in Fig.~\ref{fig:RAprotocol} and Fig.~\ref{fig:TangleProtocol}, respectively.  


The RA protocol is tailored to the agents involved and conforms to the specification in the TCG Trusted Attestation Protocol (TAP) information model~\cite{TAP}.
The first interactions (\textit{Steps 0--4}) include the provisioning phase discussed in Section~\ref{sect:assumptions}, the attestation request from $A$ to $V$, the generation of the nonce (\ie a random challenge to counteract replay attacks) by $V$, and the generation of the TPM Quote and ML to be sent to $V$. 
Then, $V$ checks the integrity of $A$ (\textit{Step 5}) using the received Quote and the ML. 
Specifically, $V$ checks the signature on the TPM Quote with $p^A_k$, then reconstructs the PCR10 using the measures in the ML and compares it to the actual PCR10 value in the Quote. Finally, $V$ checks the measures in the ML against the golden values in a whitelist. 
If any of these checks fail, $V$ terminates the communication with $A$ without generating the Attestation Report (AR). 
Otherwise, 
$V$ generates and signs the AR using its private key $s^V_k$ (\textit{Step 6}). 
Finally, $V$ sends the AR to $A$ (\textit{Step 7}) and $A$ checks the authenticity of the AR using $p^V_k$ (\textit{Step 8}).

The AR includes at least the current Timestamp $T_i$ and the public key $p^A_k$, to ensure a binding between the AR and the identity of $A$.
However, different options can be considered for the content of the AR and for the periodicity of the RA protocol, as discussed in Section~\ref{sect:timestamps}.

\begin{figure}[tbp] %
    \centerline{\includegraphics[width=\columnwidth]{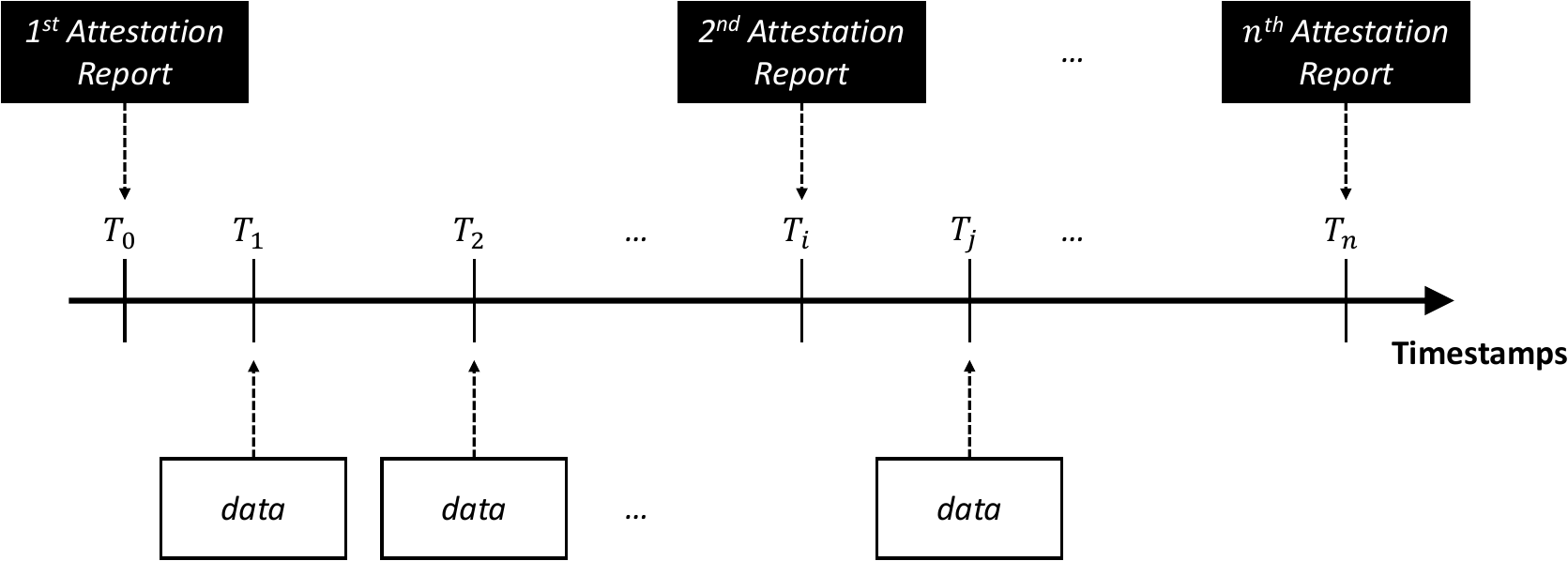}}
    \caption{Realisation of the writing process on a WAM channel.}
    \label{fig:Timestamps}
\end{figure}

Once $A$ has received an AR, the Data Exchange protocol takes place as in Fig.~\ref{fig:TangleProtocol}. In detail, $A$ writes the AR into the WAM channel (\textit{Step 9}) and, then, starts generating (\textit{Step 10}) and writing data with a valid timestamp $T_j$ into the WAM channel (\textit{Step 11}).
By periodically repeating these last three steps and the RA protocol, $A$ can actually write a continuous data stream into a WAM channel over the Tangle, consisting of a periodic succession of AR and data with proper timestamps as depicted in Fig.~\ref{fig:Timestamps}. 
The $RP$ can request access to the data from $A$ (\textit{Step 12}) at any time. $A$ provides an entry point within the WAM channel to $RP$ (\textit{Step 13}), and $RP$ uses it to directly retrieve the messages from the WAM channel of $A$ (\textit{Step 14}). 
At this point, before trusting and consuming the data, $RP$ checks (\textit{Step 15}) the authenticity of the AR, the consistency of $p_k^A$ and the timestamps in ARs.
In detail, $RP$ checks the signature on the AR using $p^V_k$
and that the public key into the AR is the expected $p_k^A$, the same public key passed to the WAM protocol to check ownership and authenticity of the messages.
Moreover, $RP$ verifies the consistency of the timestamps in data and ARs with the issuing time of the corresponding data transaction on the Tangle.

\subsection{Discussion on Key Design Aspects}  %
\label{sect:timestamps}

The agents can periodically repeat the protocols in Fig.~\ref{fig:RAprotocol} and Fig.~\ref{fig:TangleProtocol}. $A$ can periodically request a new AR from $V$ and write it into the WAM channel by executing a new iteration of the RA protocol. $A$ can generate and write fresh data into the WAM channel. These two processes are independent and $A$ can asynchronously execute them as shown in Fig.~\ref{fig:Timestamps}. At any time, $RP$ can request and obtain access to the resulting stream of data interspersed with ARs.

Different design choices result in a trade-off between security and performance of the system. In this sense, the paper identifies and briefly discusses here the main design aspects and some potential options to be considered.
\vspace{2mm} %

\textbf{AR verbosity} -- 
The AR can contain:
\begin{itemize}
    \item a synthetic report on the SW integrity of $A$, the public key $p^A_k$, the Timestamp $T_i$, and the signature computed by $V$ with $s^V_k$; 
    \item a verbose report that, in addition to the contents of the synthetic report, includes the full ML sent by $A$ to $V$ during the remote attestation.        
\end{itemize}     
The first option is preferable in near real-time use cases. The second option may be appropriate when $RP$ needs to verify the type and version of the software running at $A$. An incremental ML that contains only the new measures can be used for performance reasons. The latter requires storing the golden values at the $RP$ during the provisioning phase.

\textbf{Periodicity for checkpoints} --
The RA protocol shown in Fig.~\ref{fig:RAprotocol} and the subsequent writing of a new AR to the WAM channel can be repeated with a constant or variable time interval.
Depending on the use case, a constant time interval can be appropriate if there is a known periodicity and a predefined size for the data, while a variable interval is suitable when $A$ asynchronously generates data, possibly with a variable size that is not known \textit{a priori} (\eg 1 AR every 10 data messages or every 100 kB of data).

\textbf{Relying Party logic for trust} --  
Assuming that a $RP$ considers as trusted only the data between two valid ARs generated within a predefined threshold $TH$, the logic for the retrieval and consumption of the data can differ for non-real-time and near real-time use cases, depending on their requirements.

In a non-real-time case, $RP$ is interested in obtaining access to the full or at least a portion of the data stream, without strict latency constraints. 
Starting from an entry point on the WAM channel, which should correspond to a past AR, the $RP$ can check that the elapsed time to the next AR at is less than $TH$.
Then, the $RP$ consumes all the data between these two ARs. The same process is repeated for the subsequent ARs and data.

In a near-real-time case, the entry point on the WAM channel can be at an arbitrary time instant that probably does not coincide with an AR.
$RP$ can discard the retrieved data until the first valid AR becomes available; 
after that, $RP$ has two options for the logic of trust: 
\begin{itemize}
    \item $RP$ stores the following data in a buffer until the second AR becomes available. If and only if the two ARs are valid and the time elapsed between them does not exceed $TH$, $RP$ consumes the buffered data; 
   
    \item $RP$ retrieves and directly consumes the data without buffering. If a new AR does not become available by the expiration of $TH$, the $RP$ must reconsider the trustworthiness of the consumed data. 
\end{itemize}        
The first option can be considered more secure than the second logic that, on the other hand, seems more efficient.
In fact, the second one may reduce the latency between the writing of data by $A$ and its consumption by $RP$. 
Depending on the preferred option, $RP$ continues to process the subsequent stream of data interspersed with ARs.

\section{Experimental Results}
\label{sect:results}

\subsection{Proof-of-Concept Implementation and Configuration}
\label{sect:PoC}

The hardware infrastructure for the PoC includes:
\begin{itemize}
    \item Raspberry Pi{\textsuperscript{\textregistered}} 4 (RPi4) Model B~\cite{rpi4brief}
    (1.5 GHz CPU, 4 GB RAM, %
    Raspberry Pi OS, Linux 5.15.79-v7l+ kernel);
    \item Infineon OPTIGA{\texttrademark} TPM SLI 9670
    \cite{infineonTPM}, a board with TPM2.0 chip connected as expansion hat on RPi4 and configured to use the ECC NIST P256 curve \cite{TCGalgoRegistry, infineonTPM};
    \item mid-range server (1.8 GHz CPU, 24 GB RAM, 1 TB SSD, 
    Ubuntu 22.04 LTS, Linux 5.19.0-50 kernel);
    \item IOTA node \cite{HORNET} synchronised with the IOTA distributed ledger deployed and operated in our lab.       
\end{itemize}

The analyses consider the RPi4 with the TPM as the main device under test for both Attester and Relying Party roles.
Furthermore, the server is used as the Verifier of the IoT system, and the IOTA node is leveraged as the gateway toward the IOTA network to write and read data.
The PoC takes advantage of the RA protocol and WAM protocol implementations, both developed in C language by our research group. 

\subsection{Remote Attestation Protocol}
\label{sect:resultsRA}

An initial experimental analysis is carried out to investigate the expected performance of 
the RA protocol depicted in Fig.~\ref{fig:RAprotocol}, focusing on the operations with the largest computational complexity.

For these initial tests, the server is configured as the Verifier and two RPi4s are deployed, one as the Attester and one as the Relying Party. 
The initial provisioning phase (\textit{Step 0}) mainly consists in the generation of a full database of golden values for all the relevant files on the devices under test in a protected environment.
This phase is executed on the RPi4s and it is typically done only once for each device, thus its complexity will be excluded from the performance assessment. 

The size of the ML (\ie IMA log) can be considered as a key parameter, since it is expected to have a remarkable impact on the execution time of the main operations for the RA procedure (\textit{Steps 1--5}). 
The following analyses consider a ML size of $132$ kB,
that corresponds to approximately 900 entries on the IMA log. 
This value has been measured on the PoC just after a reboot 
and it is representative of the typical length of an IMA log on a RPi4. 
The experimental test consists in the execution of multiple independent repetitions (\ie 500 trials) of the main RA operations (\textit{Steps 1--5}) to assess the achievable performance. The Verifier measures the execution time for each trial by means of three intermediate timers:
\begin{description}
    \item[\textit{Timer1}] -- measuring the elapsed time from the reception of an attestation request from the Attester (\textit{Step 1}) up to the complete reception of the Quote and ML (\textit{Step 4});
    \item[\textit{Timer2}] -- estimating the execution time for the verification of the Quote received from Attester (first part of \textit{Step 5}); 
    \item[\textit{Timer3}] -- assessing the time to verify the ML against golden values and whitelist databases, plus the time to reconstruct and verify the PCR10 (second part of \textit{Step 5}). 
\end{description}

It is possible to compute the total RA latency as the sum of these three timers. The \textit{Steps 6--8} mainly consist in a signature generation for the AR and its verification and, thus, they are  excluded due to the limited impact on the overall performance.

\begin{table}[tbp] %
\caption{Statistical analysis of the execution time for RA operations.} %
\begin{center}
\begin{tabular}{|l|c|c|c|c|}
    \hline       
    \textbf{Statistics} & \textbf{\textit{Timer1}} (s) & \textbf{\textit{Timer2}} (s) & \textbf{\textit{Timer3}} (s) & \textbf{\textit{Total}} (s)\\
    \hline     
    \textit{max}     & 0.459 & 0.023 & 0.075 & 0.494\\
    \textit{avg}     & 0.404 & 0.009 & 0.035 & 0.447\\
    \textit{min}     & 0.358 & 0.005 & 0.021 & 0.392\\
    \textit{std}     & 0.015 & 0.003 & 0.010 & 0.018\\
    \hline    
\end{tabular}
\label{tab:RA_stat}
\end{center}
\end{table}
\begin{figure}[tbp] %
    \begin{center}
    \footnotesize %

    \includegraphics[height=6.3cm]{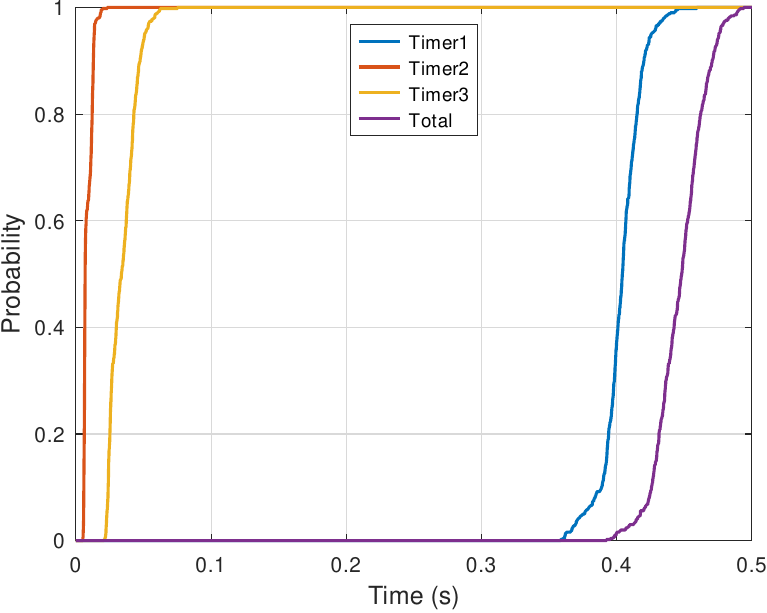}
    
    \caption{%
    Empirical CDF of the execution time for RA operations.}
    \label{fig:RA_CDF}
    \end{center}
\end{figure}

Table~\ref{tab:RA_stat} presents a statistical analysis on the experimental results, including the maximum (\textit{max}), average (\textit{avg}), minimum (\textit{min}), and standard deviation (\textit{std}) values estimated for the three timers and for the total RA time.
The same results are also depicted in Fig.~\ref{fig:RA_CDF}, that presents the empirical Cumulative Distribution Function (CDF) computed on the three intermediate timers and on the total RA time. The results in Table~\ref{tab:RA_stat} and Fig.~\ref{fig:RA_CDF} demonstrate that the \textit{Timer1}, with an average of 0.404 s, represents the largest contribution to the total RA time 
(\ie approx. 90\% of its average, that is 0.447 s). 
In fact, the \textit{Timer1} mainly assesses specific operations (\ie Quote generation, %
at \textit{Step 3}, and transmission of the Quote and ML to the Verifier, at \textit{Step 4})
that are executed on the RPi4 equipped with TPM.
On the other hand, \textit{Timer2} and \textit{Timer3} correspond to computations done by the Verifier (\textit{Step 5}). An additional investigation on the specific operations measured by \textit{Timer1} allows noticing that the Quote generation on the TPM
is the operation with the largest contribution to \textit{Timer1}, showing an average execution time of 0.361 s. It corresponds to approx. 89\% of the average value of \textit{Timer1} and to approx. 81\% of the total RA time.  
This value is justified by the fact that the Quote generation is securely executed by the TPM. 

These results confirm the feasibility of the proposed RA protocol in Fig.~\ref{fig:RAprotocol}. In fact, the measured total RA time is always below 0.5 s (\ie average equal to 0.447 s and maximum value of 0.494 s).

It is also worth to point out that the selected PoC configuration corresponds to an approximate ML size of $132$ kB. In principle, better performance (\ie lower total RA time) can be achieved by adopting a hybrid solution, where the RPi4 periodically sends a differential ML instead of transmitting the full ML. 

As an additional remark, the \textit{Timer2} is expected to have an execution time that is almost constant varying the ML size, since the assessed operation (\ie Quote verification) does not depend on the ML size, while both the \textit{Timer1} and \textit{Timer3} are significantly affected by the ML size.

\begin{figure}[tbp] %
    \footnotesize %
    \includegraphics[height=6.3cm]{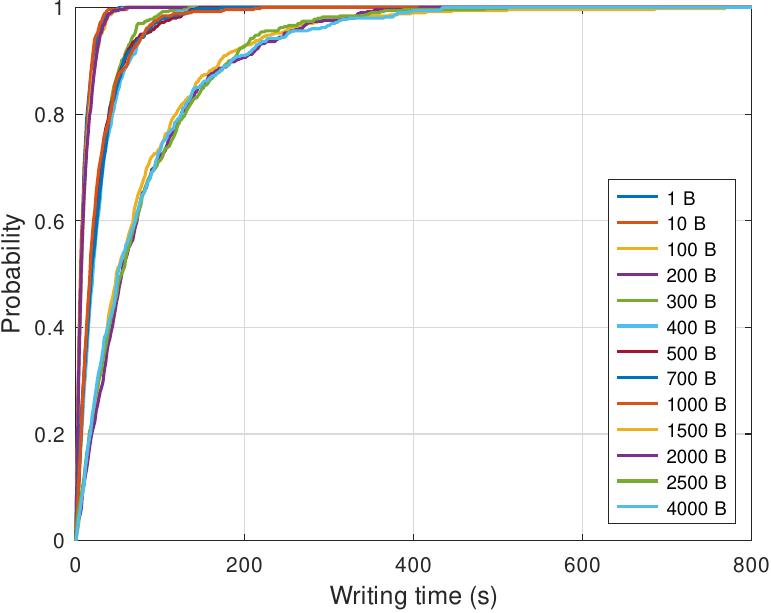}

    \vspace{2mm}
    
    \includegraphics[height=6.3cm]{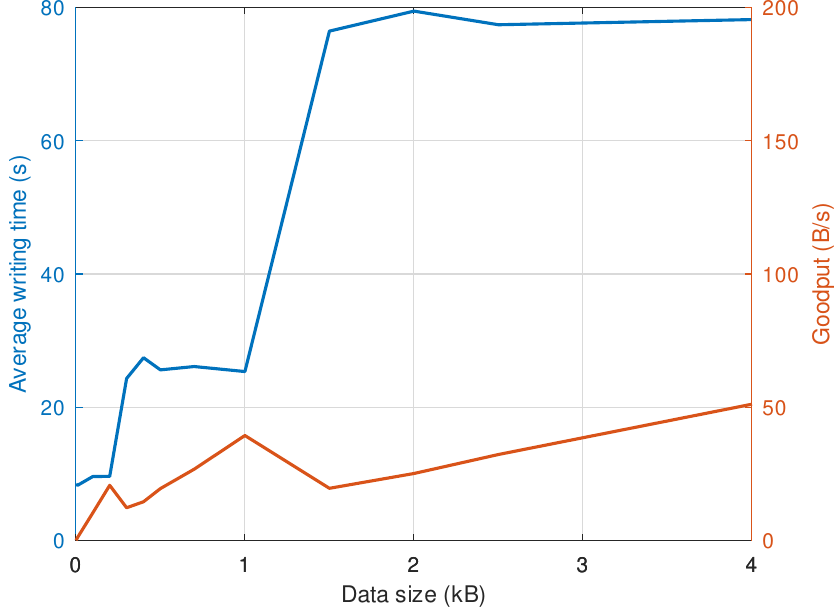}
    
    \caption{Empirical CDF of the writing time (top) and estimated average writing time and goodput (bottom) over the Tangle.}
    \label{fig:W_results}
    \vspace{-4mm}
\end{figure}
\begin{figure}[tbp] %
    \footnotesize %
    ~ %
    \includegraphics[height=6.3cm]{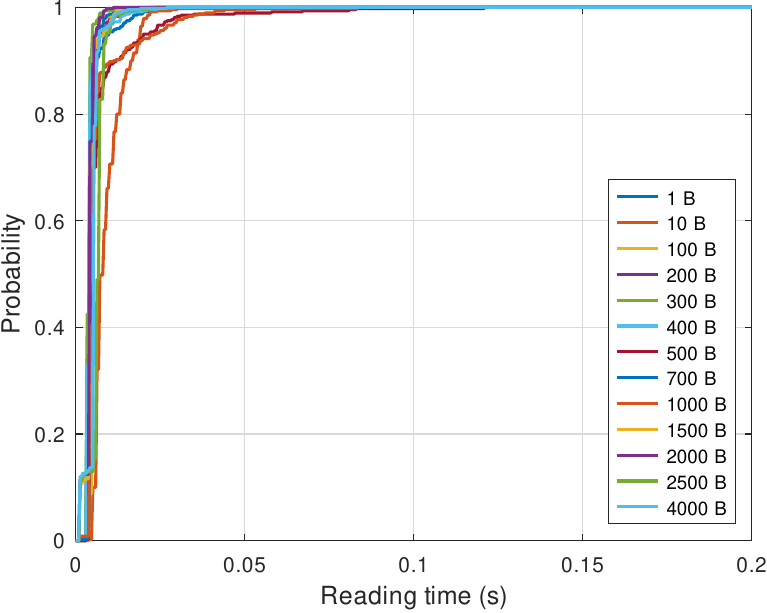}
    
    \vspace{2mm}
    
    \includegraphics[height=6.3cm]{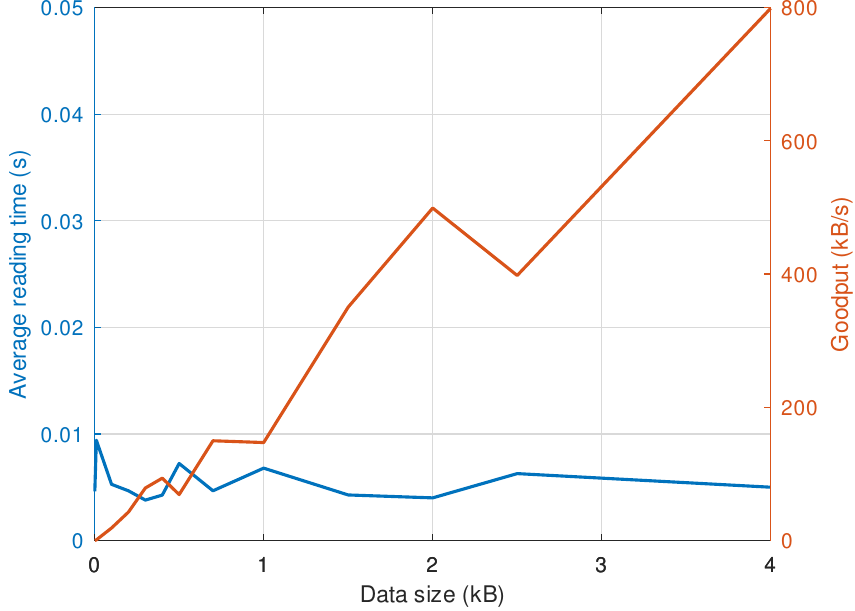}
    
    \caption{Empirical CDF of the reading time (top) and estimated average reading time and goodput (bottom) over the Tangle.}
    \label{fig:R_results}
    \vspace{-4mm}
\end{figure}

\subsection{Data Exchange Protocol}  
\label{sect:resultsWAM}

Another experimental analysis is carried out to investigate the performance of the protocol in Fig.~\ref{fig:TangleProtocol}, especially focusing on time-consuming operations on the Tangle through the WAM protocol.
In detail, it is possible to identify the following operations as the ones with the highest expected execution time:
the \textit{Steps 9} and \textit{11}, that concern the writing of specific data (\ie AR and data, respectively) into the WAM channel on the Tangle, 
and the \textit{Step 14}, that consists in retrieving such data from the WAM channel.
In fact, the remaining operations in Fig.~\ref{fig:TangleProtocol} mainly consist in data generation, timestamping, %
and signature verification, with a negligible impact on the overall complexity of the workflow and, thus, they are excluded from the following discussion.

In these tests, one RPi4 is connected to the IOTA gateway node and it emulates the periodic writing of AR and data into a WAM channel on the Tangle; it generates random data with a size from 1 byte up to 4 kilobytes. Another RPi4 retrieves the data from the WAM channel.
Such write and read operations are repeated for each selected data size and for 500 trials. 
The RPi4 nodes measure the execution time of the 
operations and these values are analysed to estimate the empirical CDF and the average execution times. 
Moreover, the analysis also includes an estimation of the Goodput $G$, that represents the number of useful information bits delivered by the network to a certain destination per unit of time.
It corresponds to the data throughput neglecting any packet overhead and can be computed as $G = D/T$, where $D$ is the considered payload data size and $T$ is the average execution time of the write or read operations.

Fig.~\ref{fig:W_results} and Fig.~\ref{fig:R_results} present the results for writing and reading data through the WAM channel on the Tangle, respectively.
In detail, Fig.~\ref{fig:W_results} (top) shows the empirical CDF of the writing time for the considered data sizes.
It can be appreciated that the CDFs are clustered in three groups, with different but consistent statistics in each group.
This behaviour is also noticeable in Fig.~\ref{fig:W_results} (bottom), that shows the estimated average writing time and the goodput versus the data size.
Each group shows an almost constant average writing time,
with the first group corresponding to the data sizes from 1 B to 200 B (and writing times from 8.3 s to 9.6 s),
the second group from 300 B to 1 kB (and writing times from 24.3 s to  27.4 s),
and the third group from 1.5 kB to 4 kB (and writing times from 76.4 s to 79.4 s).
This behaviour is well explained considering that the complexity of the Proof-of-Work that the IOTA gateway node must perform for each data transaction depends on the data size \cite{HORNET}.

The staircase pattern of the average writing time in Fig.~\ref{fig:W_results} (bottom) directly translates into a saw-tooth pattern for the goodput.
In fact, if the data size increases inside one of the groups, the almost constant average writing time implies a growing goodput.
It can be noticed that specific use cases, requiring to maximise the goodput, can be interested in finding a locally optimal value for the data size according to this pattern (\eg a WAM message payload of 1 kB results in $G = 39.4$ B/s, while 1.5 kB would imply $G = 19.6$ B/s).

A similar analysis is also done for the reading time and the results are reported in Fig.~\ref{fig:R_results}.
In this case, the measured values result in a similar empirical CDF for all data sizes considered, as shown in Fig.~\ref{fig:R_results} (top).
Thus, Fig.~\ref{fig:R_results} (bottom) presents an almost constant average reading time, with small magnitudes (from 3.8 ms to 9.4 ms).
These values result in an increasing trend for the goodput, achieving the remarkable value of $G = 798.67$ kB/s for a data size of 4 kB.    

These results confirm the feasibility of the proposed protocol in Fig.~\ref{fig:TangleProtocol}.
As expected, the achievable performance is limited for writing data to the Tangle, with average goodput values in the range of few tens of \textit{bytes} per second, while several hundreds of \textit{kilobytes} per second are achievable for reading data from the Tangle, depending on the size of each data packet.
In this sense, it seems inadvisable to periodically write a verbose AR (including the full ML) to the Tangle, unless strictly necessary, whereas it may be appropriate to use a synthetic AR. 

Moreover, if the use case requires a low latency for the delivery of data from the Attester to the Relying Party, a small packet size (\eg less than 200 B) is preferable to minimise the writing time (\eg less than 10 s). On the other hand, if the goodput must be maximised, large packet sizes (\eg from 1 kB to 4 kB) may the preferred option (\eg to achieve up to $G = 51.2$ B/s for writing and up to $G = 798.67$ kB/s for reading data).

\section{Conclusions and Future Work}
\label{sect:conclusion}

This paper has proposed TDT as a novel solution for providing the Trust in Data for IoT connected systems with stringent security requirements. The PoC implementation and the experimental results have confirmed the feasibility of the proposed solution and have showed promising performance for its main building blocks. 
Future work will focus on prototyping the TDT solution, tailoring its design parameters to real use case requirements, and extending the performance analysis to microcontroller-based devices.

\bibliographystyle{./IEEEtran}
\bibliography{Biblio_NoURL}

\end{document}